\documentclass[epj]{svjour}
\usepackage{graphics}
\usepackage{amsmath}
\newcommand{\bv}[1]{{\boldsymbol #1}}

\begin{document}
\title{Unified description of long-time tails and long-range correlation functions for sheared granular liquids }
\author{Michio Otsuki\inst{1} \and Hisao Hayakawa\inst{2}
}                     
%
%
\authorrunning{M. Otsuki and H. Hayakawa}
\titlerunning{Unified description of long-time tails and long-range correlation functions}
\institute{Department of Physics and Mathematics, Aoyama Gakuin University,
5-10-1 Fuchinobe, Sagamihara, Kanagawa, 229-8558, Japan \and Yukawa Institute for Theoretical Physics, Kyoto University,  Kitashirakawaoiwake-cho, Sakyo-ku, Kyoto 606-8502, Japan.}
\date{Received: date / Revised version: date}
%
\abstract{
Unified description on the long-time tail of velocity autocorrelation function (VACF) and the algebraic decays of the equal-time spatial correlation functions for nearly elastic and uniformly sheared granular liquids is developed
based on the generalized fluctuating hydrodynamics. We predict that
the cross-over of the long-time tail of VACF from $t^{-3/2}$ to $t^{-5/2}$ with the time $t$ regardless of the density.
We also demonstrate the existence of algebraic tails of
the equal-time spatial density correlation function and the equal-time spatial velocity correlation function which respectively 
satisfy $r^{-11/3}$ and $r^{-5/3}$ for large distance $r$.   
} 

\maketitle
\section{Introduction}
\label{Intro:sec}

 Many theoretical aspects of granular assemblies have been described by gas kinetic theory such as Boltzmann-Enskog equation \cite{brilliantov04,goldhirsch} in which effect of correlations only appears through the radial distribution function at contact.
Based on Chapman-Enskog method or Grad expansion, it is possible to determine the transport coefficients and 
to derive hydrodynamic equations \cite{JR,Brey98,Garzo,Lutsko05}.
In spite of the success of semi-quantitative description of granular assemblies by the kinetic theory \cite{Saitoh}, 
recently we have recognized important roles of long-time tails and long-range correlation functions 
in granular liquids \cite{Kumaran06,Kumaran09,Orpe07,Orpe08,Hayakawa07,otsuki07,Hayakawa09s,Otsuki09s,Otsuki09r,Otsuki09,Rycroft} 
which
cannot be described by Boltzmann-Enskog theory.
 
In general, correlation effects are complicated, which strongly depend on the boundary condition.
However,  we can extract simple statistical properties of 
uniformly sheared granular liquids which are much simpler than nonuniformly sheared systems and 
independent of the boundary condition. 
Indeed, recent analysis on uniformly sheared granular liquids with small inelasticity 
has revealed the existence of long-range spatial correlation function,
the suppress of the long-time tail of velocity autocorrelation function, the existence of generalized Green-Kubo formula,
 and the integral fluctuation theorem \cite{otsuki07,Otsuki09,Chong}.
This uniformly sheared granular system can be achieved, at least, even for boundary-driven 
dense flow of frictionless granular particles \cite{Hatano07}.
Moreover, we need to remove the boundary effects to extract the spatial correlation function of granular liquids
which should be translational invariant.
However, our previous analysis on the long-time tails assume that granular gas is dilute, while experiments analyze the dense flows. 
So far, the connection between  the long-time tails for velocity autocorrelation function and the equal-time spatial correlation functions is not clear.
Hence, we demonstrate both the long-time tails and the long-range correlation functions 
for uniformly sheared granular liquids with small inelasicity can be derived by an unified method 
based on the generalized fluctuating hydrodynamics \cite{Kirkpatrick85,Kirkpatrick86,Das90,Marchetti}.

In this paper, thus, we theoretically investigate 
the velocity autocorrelation function (VACF) and the equal-time correlation functions  based on the generalized fluctuating
hydrodynamics with a linear non-local but instanteneous
constitutive relation and fluctuation-dissipation relation. 
As stressed in Refs. \cite{otsuki07,Otsuki09,Lutsko85}, 
the formulation can be used for any sheared isothermal liquids besides granular liquids.
However, to clarify the argument, we only focus on uniformly sheared granular liquids with small inelasticity.
In Sec. \ref{GFH:sec}, we will 
summarize the outline of the generalized fluctuating hydrodynamics.
In Sec. \ref{Lin:sec} we will analyze the linearized  generalized fluctuating hydrodynamics.
In Sec. \ref{Cor:sec}, we will calculate the correlation functions
and their asymptotic forms in the long-time limit and the large distance limit.
In Sec. \ref{Dis:sec} we will discuss and conclude our results.
In Appendices, we present some details of our calculations.

\section{Generalized fluctuating hydrodynamics}
\label{GFH:sec}

The systems we consider are three-dimensional sheared granular flows consisting of $N$ 
identical smooth and inelastic hard spherical particles in the volume $V$,
where the mass and the diameter of each grain are respectively given by
$m$ and $\sigma$. 
We should note that granular systems often cause shear-bands near the boundary. 
Since we are not interested in complicated behaviors under such shear bands, 
we focus on the uniformly sheared granular liquids characterized by 
$\alpha$ component of the velocity field $c_\alpha(\bv{r}) = \dot \gamma y \delta_{\alpha,x}$ with the shear rate $\dot\gamma$.
It is known that such uniform flow can be realized if the system size is not extremely large and granular assemblies are nearly elastic  under Lees-Edwards boundary condition.  
We also assume that the particles collide instantaneously with each other
by a restitution constant $e$ which is less than unity.

Let us consider the velocity autocorrelation function in the 
three-dimensional sheared granular liquids,
defined by
\begin{eqnarray}
C_{{\rm S}, \alpha \alpha}(t) & \equiv  & \frac{1}{N} \sum_i^N
\langle  \delta v_{i,\alpha}(t)  \delta v_{i,\alpha}(0) 
\rangle, 
\label{C:def}
\end{eqnarray}
where 
$\delta \bv{v}_i(t) \equiv \bv{v}_i(t) - \bv{c}(\bv{r}_i(t))$ 
is the peculiar velocity of the particle $i$.
In addition, we also consider  the equal-time spatial correlation functions under the steady state
\begin{eqnarray}
C_{nn}(\bv{r}) & \equiv & \langle  \delta n(\bv{r}+\bv{r}',0) 
\delta n (\bv{r}',0)\rangle,  \label{Cnn:def}\\
C_{pp}(\bv{r}) & \equiv & \langle  \delta \bv{p}(\bv{r}+\bv{r}',0) 
\cdot
\delta \bv{p} (\bv{r}',0)\rangle \label{Cpp:def}
\end{eqnarray}
for the fluctuations of the number density $\delta n(\bv{r},t)\equiv n(\bv{r},t)-n_0$ and
the momentum density $\delta \bv{p}(\bv{r},t) \equiv
 m \ n(\bv{r},t) \delta \bv{u}(\bv{r},t)$ with
 the fluctuation of the velocity field $\bv{u}(\bv{r},t)$ given
 as $\delta \bv{u}(\bv{r},t) = \bv{u}(\bv{r},t) - \bv{c}(\bv{r})$.

Introducing the Fourier transform of any function 
$\hat{f}(\bv{q}) = \int d \bv{r} f(\bv{r}) e^{i \bv{q} \cdot \bv{r}}$,
and following Ref. \cite{Zwanzig},
we can rewrite $C_{{\rm S}, \alpha \alpha}(t) $  as
\begin{eqnarray}
C_{{\rm S}, \alpha \alpha}(t) & =  & 
\int \frac{d \bv{q}}{(2\pi)^3} \frac{d \bv{q}'}{(2\pi)^3} 
\langle  \delta 
\hat{u}_\alpha(\bv{q},t) \delta 
\hat{u}_\alpha(\bv{q}',0) \rangle \nonumber \\
& &
\langle  
\hat{P}(\bv{q},t) 
\hat{P}(\bv{q}',0) 
\rangle,
\label{CS:eq}
\end{eqnarray}
where $\hat{P}(\bv{q},t)$ is
the Fourier transform of 
the the microscopic {\it concentration} of the tagged particle $i$
defined by $P(\bv{r},t) \equiv \delta(\bv{r}_i(t) - \bv{r} )$ (see Appendix
 \ref{CS:app}).

Similarly, the spatial correlations in Eqs. (\ref{Cnn:def}) and (\ref{Cpp:def})
can be rewritten as
\begin{eqnarray}
C_{nn}(\bv{r}) & = & 
\int \frac{d\bv{q}}{(2\pi)^3} \frac{d\bv{q}'}{(2\pi)^3} 
\langle  \delta \hat{n}(\bv{q},0) 
\delta \hat{n} (\bv{q}',0)\rangle \nonumber \\
& & e^{-i\bv{q} \cdot \bv{r}-i(\bv{q}+\bv{q}') \cdot \bv{r}'},  \label{Cnn:rew}\\
C_{pp}(\bv{r}) & \simeq & (m \ n_0)^2 
\int \frac{d\bv{q}}{(2\pi)^3} \frac{d\bv{q}'}{(2\pi)^3} 
\langle  \delta \hat{\bv{u}}(\bv{q},0) 
\cdot
\delta \hat{\bv{u}} (\bv{q}',0) \rangle \nonumber \\
& & e^{-i\bv{q} \cdot \bv{r}-i(\bv{q}+\bv{q}') \cdot \bv{r}'}, 
\label{Cpp:rew}
\end{eqnarray}
where we have used the Fourier transform of $\delta n(\bv{r},t)$ and
$\delta \bv{p}(\bv{r},t)$, and 
$\delta \bv{p}(\bv{r},t) \simeq 
 m \ n_0 \delta \bv{u}(\bv{r},t)$.

Since we assume uniformly sheared case, the temperature $T$ only appears through the time evolution equation of the velocity fields $\bv{u}(\bv{r},t)$. 
Thus,  the equations for the generalized hydrodynamic equations are given by
\cite{Kirkpatrick85,Kirkpatrick86,Das90,Marchetti,Otsuki09}
\begin{equation}
\partial_t n + \bv{\nabla} \cdot (n \bv{u})  =  0, 
\label{n:eq} 
\end{equation}
\begin{equation}
\partial_t u_\alpha + u_\alpha \nabla_\beta u_\beta + \frac{1}{m} \nabla_{\alpha}
\mu
+ \frac{1}{mn} \nabla_\beta (\Sigma_{\alpha \beta}^D + \Sigma_{\alpha \beta}^R)
 = 0,  \label{u:eq} 
\end{equation}
where  $n$
is  the number density.
Here, we have introduced the generalized chemical potential or
the effective pressure
\begin{eqnarray}
\mu= T \left [ \ln n - \int d\bv{r}' C(\bv{r} - \bv{r}',e,\dot\gamma)
\delta n(\bv{r}',t) + \cdots
\right ],
\end{eqnarray}
where $C(\bv{r}-\bv{r}',e,\dot\gamma)$ is the direct correlation function, which satisfies
$nC(k,e,\dot\gamma) \equiv 1 - S(k,e,\dot\gamma)^{-1}$ with the structure factor $S(k,e,\dot\gamma)$.
$\Sigma_{\alpha \beta}^D(\bv{r},t)$ is the viscous stress tensor given by
\begin{eqnarray}\label{noise}
& &\Sigma_{\alpha \beta}^D(\bv{r},t)
=
 -\int d \bv{r}' \left [ 
\eta(\bv{r}-\bv{r}',e) 
\dot{\epsilon}_{\alpha \beta}(\bv{r}',t) \right . \nonumber \\
& & \qquad \left .
- \{ 2\eta(\bv{r}-\bv{r}',e)/3
- \zeta(\bv{r}-\bv{r}',e) \}
\dot{\epsilon}_{\gamma\gamma}(\bv{r}',t)\delta_{\alpha \beta} \right ],
\end{eqnarray}
where $\dot{\epsilon}_{\alpha \beta}(\bv{r},t)\equiv
\{ \nabla_\alpha u_\beta(\bv{r},t) + \nabla_\beta u_\alpha(\bv{r},t) \} /2$.
$\Sigma_{\alpha \beta}^R(\bv{r},t)$ is the random part of the stress tensor satisfying 
$\langle \Sigma_{\alpha\beta}^R
 \rangle=0$, and the fluctuation-dissipation relation (FDR)
\begin{eqnarray}\label{noise-corr}
\langle \Sigma_{\alpha\beta}^R(\bv{r},t)  \Sigma_{\gamma\delta}^R(\bv{r}',t')\rangle& = &
2T \delta(t-t')  
 \{ \eta(\bv{r}-\bv{r}',e) \Delta_{\alpha\beta\gamma\delta} \nonumber \\
& & + \zeta(\bv{r}-\bv{r}',e) 
\delta_{\alpha\beta} \delta_{\gamma\delta} \}
\end{eqnarray}
 with $\Delta_{\alpha\beta\delta\gamma}
 \equiv \delta_{\alpha\gamma}\delta_{\beta\delta} + 
\delta_{\alpha\delta} \delta_{\beta\gamma} - 2\delta_{\alpha\beta}\delta_{\gamma\delta}/3$.
Here, we have used Einstein's sum rule on the Greek subscript. 
The generalized shear viscosity $\eta(\bv{r},e)$ and the generalized bulk viscosity $\zeta(\bv{r},e)$ are
represented by $\nu^*_1(k,e)$, $\nu^*_2(k,e)$ and Enskog's mean free time 
\begin{equation}\label{enskog_time}
t_E\equiv \frac{1}{4\pi n_0\sigma^2 g_0(\sigma,e)}
\left(
\frac{m\pi}{T} 
\right)^{1/2}  
\end{equation}
  with
the radial distribution function $g_0(\sigma,e)$
at contact
 as  $\nu^*_1(k,e) = (mn_0 \sigma^2 t_E^{-1})^{-1}
(\zeta(k,e) + 4\eta(k,e)/3)$, and $\nu^*_2(k,e)= (mn_0 \sigma^2 t_E^{-1})^{-1}
\eta(k,e)$, where $n_0$, $\eta(k,e)$ and $\zeta(k,e)$ are respectively the average number density, and Fourier transforms of $\eta(\bv{r},e)$ and
$\zeta(\bv{r},e)$.

It is known that $\nu^*_1(k,e)$ and $\nu^*_2(k,e)$
are respectively given by 
\begin{equation}\label{nu1}
\nu^*_1(k,1) = 2(1-j_0(k)+2j_2(k))/(3k^2) ,
\end{equation}
and 
\begin{equation}\label{nu2}
\nu^*_2(k,1) = 2(1-j_0(k)-j_2(k))/(3k^2)
\end{equation}
for elastic hard spherical particles, 
where $j_l(k)$ with $l=0$ or $2$ is the $l$-th. order spherical Bessel function \cite{Kirkpatrick85,Kirkpatrick86,Das90,Marchetti,deSchepper,Alley83}.
Although we do not know how $\nu^*_1(k,e)$ and $\nu^*_2(k,e)$ depend on $e$, the explicit $e$-dependences
of $\nu^*_1(k,e)$ and $\nu^*_2(k,e)$ are not
important in this paper. Therefore, we are keeping discussion without their explicit forms.

It should be noted that 
the equilibrium or the unsheared structure factor 
$S_0(k,e)\equiv S(k,e,\dot\gamma=0)$ is given by 
an approximate expression of the pair-correlation function 
for unsheared granular liquids \cite{lutsko01},
which covers the equilibrium pair-correlation in the elastic limit.
In Eq. \eqref{noise}, we assume a linear and instantaneous
transport law,
which is valid when the inelasticity of the particles
is small in the granular liquid.
If the inelasticity is not small enough, we need to use
a non-linear transport laws because the normal stress differs from the shear stress \cite{Goldhirsh}.
This set of equations \eqref{n:eq}-\eqref{noise-corr} 
is a reasonable starting point for isothermal molecular liquids ($e=1$), 
once we use appropriate generalized
transport coefficients and $S_0(k,e=1)$,
and can be used to describe isothermal sheared granular fluids in the vicinity of $e=1$ \cite{Otsuki09}.
We also note that FDR (\ref{noise-corr}) is not satisfied in highly dissipative granular liquids. 
Indeed, we can demonstrate
the existence of some corrections in the expression of the noise correlation for general case. 
It is possible to {\it derive} the fluctuating hydrodynamics from the Liouville equation and its corresponding Mori equation
with the Markovian approximation and dropping some correlated noise terms 
\cite{Hayakawa09}.
The use of FDR, however, can be justified if the system is nearly elastic and weakly sheared situation and the density is far from
jamming transition point (point J).
The validity of the generalized fluctuation hydrodynamics has been verified through the comparison between the theory and the simulation
if the density is far from point J \cite{otsuki07,Otsuki09}. We expect that non-Markovian effects play important roles 
near point J.

\section{Linearized equations around uniform shear flow and their solutions}
\label{Lin:sec}

In this section, we analyze the hydrodynamic fluctuations
given by Eqs. \eqref{n:eq} and \eqref{u:eq}.
In the first part, we introduce the linearized equations
for the fluctuations.
In the second part, we explicitly write the solution of the linearized equations.

\subsection{Linearized equations}

Let us introduce 
  the non-dimensionalized vector $\bv{z}(\bv{r},t)$, whose  Fourier transform is 
  given by
\begin{eqnarray}
\hat{\bv{z}}^T({\bv{q}},t) & = & ( {\delta \hat{n}}(\bv{q},t), 
 {\delta \hat{u}}_x(\bv{q},t)/(t_E^{-1} \sigma^4), \nonumber \\
 & & {\delta \hat{u}}_y(\bv{q},t)/(t_E^{-1} \sigma^4), 
 { \delta \hat{u}}_z(\bv{q},t)/(t_E^{-1} \sigma^4)). \label{z:def}
\end{eqnarray}
From Eqs. (\ref{n:eq}), (\ref{u:eq}), and (\ref{z:def}), we obtain  the linearized evolution equation for $\tilde{\bv{z}}(\bv{k},\bar{t})\equiv \hat{\bv{z}}(\bv{q},t)$ as
\begin{equation}
\left(\partial_{\bar{t}} - \dot{\gamma}^* k_x \frac{\partial}{\partial k_y} \right) \tilde{\bv{z}}
+ {\sf L}\cdot \tilde{\bv{z}} = \tilde{\bv{R}},
 \label{Lin:eq}
 \end{equation}
 where 
 the time, the wavenumber and the shear rate have been non-dimensionalized  by
 $t= t_E \bar{t}$, $\bv{q} = \bv{k}/\sigma$,
 and $\dot{\gamma} = \dot{\gamma}^*/t_E$,
 respectively.  
Here, the matrix ${\sf L}$ is expanded as 
\begin{equation}
{\sf L} =  {\sf L}_{0} + \dot{\gamma}^* {\sf L}_{1} + \cdots, \label{matrix}
  \end{equation}
where ${\sf L}_{0}$ and ${\sf L}_{1}$ are respectively given by
\begin{equation}
{\sf L}_{0} =
-{\sf L}_{S0} + {\sf L}_{L0} + {\sf L}_{D0}
  \end{equation}
with
\begin{eqnarray}
{\sf L}_{S0} & = & 
\left[ 
\begin{array}{cccc}
0 &  n_0 \sigma^3 ik_x &  n_0 \sigma^3 ik_y &  n_0 \sigma^3 ik_z \\
 p^* ik_x  & 0 &
0 & 0  \\
 p^* ik_y & 0 & 0 & 
0  \\
 p^* ik_z & 0 & 0 & 
0 \\
\end{array} 
\right],
\end{eqnarray}
\begin{eqnarray}
{\sf L}_{L0} & = & 
\left[ 
\begin{array}{cccc}
0 & 0 & 0 & 0 \\
0  & \nu^*_3 k_x^2 &
 \nu^*_3 k_xk_y & \nu^*_3 k_x k_z  \\
0 & \nu^*_3 k_y k_x & \nu^*_3 k_y^2 & 
\nu^*_3 k_y k_z  \\
0 & \nu^*_3 k_z k_x & \nu^*_3 k_z k_y & 
\nu^*_3 k_z^2 \\
\end{array} 
\right],
\end{eqnarray}
\begin{eqnarray}
{\sf L}_{D0} & = & 
\left[ 
\begin{array}{cccc}
0 & 0 & 0 & 0 \\
0  &  \nu^*_2k^2 &
0 & 0  \\
0 & 0 & \nu^*_2 k^2 & 
0  \\
0 & 0 &  & 
\nu^*_2 k^2 \\
\end{array} 
\right],
\end{eqnarray}
and
\begin{eqnarray}
{\sf L}_{1} & = & 
\left[ 
\begin{array}{cccc}
0 & 0 & 0 & 0  \\
0  & 0 & 1 & 0  \\
0  & 0  & 0 & 0  \\
0  & 0 & 0 & 0  \\
\end{array} 
\right]
\end{eqnarray}
with $\nu^*_3 \equiv \nu^*_1-\nu^*_2$, $p^*_1\equiv p^*(k,e) = A S_0(k,e)^{-1}$ 
and $A \equiv T /(mn_0 \sigma^5 t_E^{-2} )$.
The random vector $\tilde{\bv{R}}$  has four components 
\begin{eqnarray}\label{R_a}
\tilde{R}_1 & = & 0, \nonumber \\
\tilde{R}_{\alpha+1} & = & (mn_0 \sigma^4 t_E^{-2})^{-1} i \sigma k_\beta \Sigma_{\alpha\beta}^R(\bv{k},\bar{t}),
\end{eqnarray}
where $\alpha=1,2,3$ respectively correspond to $x, y$ and $z$.
Although $\dot\gamma\*$ dependence of $S(k,e,\dot\gamma)$ should appear in ${\sf L}_1$,
we simply ignore such terms. The validity of this simplification has already been checked from the comparison of the results with our simulation \cite{Otsuki09}.

We should note that the wave number $k_\alpha$ is always larger than
$\Lambda=2 \pi /L$ with the system size $L$,
because the uniformly sheared flow becomes unstable for sheared granular liquids
in the limit $\bv{k} \rightarrow 0$, 
and we need infinitely large work to produce the uniformly 
shear flow for all situations.
This singular behavior around $k_x=0$ can be understood
because the effect of the shear in the left hand side of Eq. (\ref{Lin:eq})
disappears if we take $k_x=0$.
Thus, it is not appropriate to take the limit $k_x\to 0$ without the introduction of the infrared cutoff $\Lambda$.

\subsection{Solution of the linearized equations}

The solution of the linearized equation \eqref{Lin:eq} is
easily obtained following the parallel procedure in \cite{Otsuki09,Lutsko85}. 
We introduce
the right eigenvector 
$\bv{\psi}^{(j)}(\bv{k})$ and the eigenvalue $\lambda^{(j)}(\bv{k})$ satisfying
\begin{equation}
\left(- {\sf 1}\dot{\gamma}^* k_x \frac{\partial}{\partial k_y}  + {\sf L} \right)\cdot
\bv{\psi}^{(j)}(\bv{k}) = \lambda^{(j)}(\bv{k}) \bv{\psi}^{(j)}(\bv{k}).
\label{eigen:eq}
\end{equation}
We also introduce the associated biorthogonal vector, {\it i.e.} the left eigenvector $\bv{\varphi}^{(j)}(\bv{k})$,
satisfying
$\bv{\psi}^{(i)}(\bv{k})\cdot \bv{\varphi}^{(j)}(\bv{k})= \delta_{kj}$.

Following the procedure described in Appendix \ref{eigen:app},
we obtain the eigenvalues
\begin{eqnarray}
\lambda^{(1)} & = & \lambda_+ + \dot{\gamma}^* \frac{k_xk_y}{k^2} \xi^{(1)}(k), \label{lam1} \\
\lambda^{(2)} & = & \lambda_- + \dot{\gamma}^* \frac{k_xk_y}{k^2} \xi^{(2)}(k), 
\label{lam2}\\
\lambda^{(3)} & = & \nu^*_2(k,e)k^2 - \dot{\gamma}^* \frac{k_xk_y}{k^2},  
\label{lam3}\\
\lambda^{(4)} & = & \nu^*_2(k,e)k^2 \label{lam4}
\end{eqnarray}
within the approximation up to $O(\dot\gamma^*)$, where we have introduced
\begin{equation}
\lambda_+  = \frac{\nu^*_1(k,e)k^2 + \sqrt{(\nu^*_1(k,e)k^2)^2 - 4n_0 \sigma^3p^*(k,e)k^2}}{2}, 
\end{equation}
\begin{equation}
\lambda_-  = \frac{\nu^*_1(k,e)k^2 - \sqrt{(\nu^*_1(k,e)k^2)^2 - 4n_0 \sigma^3p^*(k,e)k^2}}{2},
\end{equation}
\begin{eqnarray} 
\xi^{(1)}(k)& \equiv & \frac{\lambda_+^2}{N_+^2} + \frac{n_0\sigma^3k^2}{2N_+^2}
k \partial_k p^*(k,e), \\
\xi^{(2)}(k) & \equiv &\frac{\lambda_-^2}{N_-^2} + \frac{n_0\sigma^3k^2}{2N_-^2}
k \partial_k p^*(k,e),
\end{eqnarray}
and
\begin{eqnarray}
N_+^2 & = & - n_0 \sigma^3p^*(k,e)k^2 + \lambda_+^2,  \\
N_-^2 & = & - n_0 \sigma^3p^*(k,e)k^2 + \lambda_-^2.
\end{eqnarray}
See Appendix \ref{eigen:app} for the detailed expressions of the eigenvectors.

Using the eigenvectors $\bv{\psi}^{(j)}(\bv{k})$ and $\bv{\varphi}^{(j)}(\bv{k})$,
the vector $\tilde{\bv{z}}(\bv{k},\bar{t})$ is represented by
\begin{equation}
\tilde{\bv{z}}(\bv{k},\bar{t})
= \sum_j^4 \tilde{a}^{(j)}(\bv{k},\bar{t})\bv{\psi}(\bv{k}),
\label{z:re}
\end{equation}
where we have introduced
$\tilde{a}^{(j)}(\bv{k},\bar{t}) \equiv \bv{\varphi}^{(j)}(\bv{k})
\cdot \tilde{\bv{z}}(\bv{k},\bar{t})$.
Substituting Eq. \eqref{z:re} into Eq. \eqref{Lin:eq} and
taking the inner product with $\bv{\psi}^{(j)}(\bv{k})$,
we obtain the time evolution equation of $\tilde{a}^{(j)}(\bv{k},\bar{t})$
as
\begin{equation}
\left(\partial_{\bar{t}} - \dot{\gamma}^* k_x \frac{\partial}{\partial k_y} 
+ \lambda^{(j)}(\bv{k}) \right) \tilde{a}^{(j)}(\bv{k},\bar{t})
 = F^{(j)}(\bv{k},\bar{t}), \label{a:eq}
 \end{equation}
where we have introduced
\begin{equation}
F^{(j)}(\bv{k},\bar{t}) \equiv \bv{\varphi}^{(j)}(\bv{k})\cdot\tilde{\bv{R}}(\bv{k},\bar{t}).
\label{F^j}
\end{equation}

As shown in Appendix \ref{sol:app}, 
the solution of Eq. \eqref{a:eq} is given by
\begin{equation}
\tilde{a}^{(j)}(\bv{k},\bar{t}) = \int _{-\infty}^{\bar{t}} ds
E^{(j)}(\bv{k},\bar{t}-s)
F^{(j)}(\tilde{\bv{k}}(\dot{\gamma}^*(\bar{t}-s)),s), \label{a:sol}
\end{equation}
where 
\begin{eqnarray}
E^{(j)}(\bv{k},\bar{t})
&\equiv&
\exp[  - \int_0^{\bar{t}} d\bar{\tau} \lambda^{(j)}( \tilde{\bv{k}}(\dot{\gamma}^*\bar{\tau}))], \label{E:def}
\end{eqnarray}
with $\tilde{\bv{k}}(\bar{\tau}) \equiv (k_x,k_y + \bar{\tau} k_x,k_z)$.
Hence, the solution of Eq. (\ref{Lin:eq}) 
can be formally represented by  
\begin{equation}
\tilde{\bv{z}}(\bv{k},\bar{t}) = 
\sum_{j=1}^4 \int _{-\infty}^{\bar{t}} ds
\tilde{\bv{\psi}}^{(j)}(\bv{k},\bar{t}-s)
F^{(j)}(\tilde{\bv{k}}(\dot{\gamma}^*(\bar{t}-s)),s), \label{z:sol}
\end{equation}
where 
\begin{eqnarray}
\tilde{\bv{\psi}}^{(j)}(\bv{k},\bar{t})
&\equiv&
\bv{\psi}^{(j)}(\bv{k}) E^{(j)}(\bv{k},\bar{t}).
\label{psit:def}
\end{eqnarray}

\section{Correlation functions}
\label{Cor:sec}

This section is the main part of this paper, 
in which we present the explicit forms of correlation functions.
This section consists of three parts. 
The first part summarizes the general results of correlation functions.
In the second part, we evaluate the long-time tail of the 
velocity autocorrelation function.
In the third part, we discuss the equal-time long-range correlation functions 
which are essentially the same as that in Ref. \cite{Otsuki09}.

\subsection{General results for correlation functions}

In order to obtain the correlations from Eq. \eqref{CS:eq}--\eqref{Cpp:rew},
we consider $\langle  \tilde{z}_\alpha(\bv{k},\bar{t}) \tilde{z}_\beta(\bv{k}',\bar{t})\rangle$, which is represented by 
\begin{eqnarray}
& &\langle  \tilde{z}_\alpha(\bv{k},\bar{t}) \tilde{z}_\beta(\bv{k}',0)\rangle
\nonumber \\
& &=
\sum_{l,m=1}^4 \int _{-\infty}^{\bar{t}} ds\int _{-\infty}^{0} ds'
\tilde{\psi}_\alpha^{(l)}(\bv{k},\bar{t}-s)\tilde{\psi}_\beta^{(m)}(\bv{k}',-s')
\nonumber \\
& &\quad
\times \langle F^{(l)}(\tilde{\bv{k}}(\dot{\gamma}^*(\bar{t}-s)),s)
F^{(m)}(\tilde{\bv{k}}'(\dot{\gamma}^*(-s')),s') \rangle.
\end{eqnarray}
From Eqs. \eqref{R_a} and \eqref{F^j},
$\langle F^{(l)}(\bv{k},\bar{t}) F^{(m)}(\bv{k}',\bar{t}')\rangle $
satisfies
\begin{equation}
\langle F^{(l)}(\bv{k},\bar{t}) F^{(m)}(\bv{k}',\bar{t}')\rangle = 
(2\pi)^3 \delta^3(\bv{k}+\bv{k}') \delta(\bar{t}-\bar{t}') F^{(lm)}(\bv{k}),
\label{Flm:eq}
\end{equation}
with
\begin{eqnarray}
F^{(11)}(\bv{k}) & = & - 2A k^2\nu^*_1(k,e) \frac{\lambda_+^2}{N_+^2},
\nonumber \\
F^{(22)}(\bv{k}) & = & - 2A k^2\nu^*_1(k,e) \frac{\lambda_-^2}{N_-^2},
\nonumber \\
F^{(12)}(\bv{k}) & = & F^{(21)}(\bv{k},\bar{t}) = - 2A k^2\nu^*_1(k,e) \frac{\lambda_+ \lambda_-}{N_+ N_-},
\nonumber \\
F^{(33)}(\bv{k}) & = &  2A k^2\nu^*_2(k,e),
\nonumber \\
F^{(44)}(\bv{k}) & = &  -(M(\bv{k})^2 + 1)F^{(33)}(\bv{k}),
\nonumber \\
F^{(34)}(\bv{k}) & = &  -F^{(43)}(\bv{k}) =  M(\bv{k}) F^{(33)}(\bv{k}),
\label{F_ij}
\end{eqnarray}
and 
\begin{eqnarray}
F^{(lm)}(\bv{k}) = 0,
\label{F:other}
\end{eqnarray}
for other pairs of $(l,m)$.
Hence, we obtain
\begin{eqnarray}
& &\langle  \tilde{z}_\alpha(\bv{k},\bar{t}) \tilde{z}_\beta(\bv{k}',0)\rangle
\nonumber \\
& &=
\sum_{l,m=1}^4 \int _{-\infty}^{\bar{t}} ds\int _{-\infty}^{0} ds'
\tilde{\psi}_\alpha^{(l)}(\bv{k},\bar{t}-s)\tilde{\psi}_\beta^{(m)}(\bv{k}',-s')
\nonumber \\
& &\quad  \times (2\pi)^3 \delta^3(\tilde{\bv{k}}(\dot{\gamma}^*(\bar{t}-s))+
\tilde{\bv{k}}'(\dot{\gamma}^*(-s'))) \nonumber \\
& & \quad \times \delta(s-s') F^{(lm)}( \tilde{\bv{k}}(\dot{\gamma}^*(\bar{t}-s')) )
\nonumber \\
&  &=
(2\pi)^3 \delta^3
(\tilde{\bv{k}}(\dot{\gamma}^*\bar{t})+ \bv{k}')  
C_{\alpha \beta}(\bv{k},t),
\label{Z:cor}
\end{eqnarray}
where we have introduced
\begin{eqnarray}
C_{\alpha \beta}(\bv{k},t)
& \equiv &
\sum_{l,m=1}^4 \int ^{\infty}_{0} d\tau
\tilde{\psi}_\alpha^{(l)}(\bv{k},\bar{t}+\tau)
\tilde{\psi}_\beta^{(m)}( -\tilde{\bv{k}}(\dot{\gamma}^*\bar{t}),\tau)
\nonumber \\
& & \times F^{(lm)}( \tilde{\bv{k}}(\dot{\gamma}^*(\bar{t}+\tau)) ). 
\label{C:ab}
\end{eqnarray}

\subsection{Long time tails}
\label{long_time:sec}

VACF has already been obtained 
for the dilute granular gases under a simpler approximation in Ref. \cite{otsuki07},
but we can discuss them by the more precise treatment mentioned in this paper.

To obtain  $C_{{\rm S}, \alpha \alpha}(t)$ in Eq. \eqref{CS:eq},
we need to solve the time evolution of 
$\tilde{P}(\bv{k},\bar{t}) \equiv \hat{P}(\bv{q},t)$
given by
\begin{equation}
\left(\partial_{\bar{t}} - \dot{\gamma}^* k_x \frac{\partial}{\partial k_y} 
+ D^* k^2 \right) \tilde{P}(\bv{k},\bar{t})
 = 0,
 \end{equation}
 with $D^* = (\sigma^{-2} t_E)D$.
The solution of this equation is given by
\begin{equation}
\tilde{P}(\bv{k},\bar{t}) = \tilde{P}(\tilde{\bv{k}}(\dot{\gamma}^*\bar{t}),0) 
E_D(\bv{k},\bar{t}),
 \end{equation}
 with
\begin{equation}
E_D(\bv{k},\bar{t}) = \exp \left [ 
- \int_0^{\bar{t}} ds D^* \tilde{k}(\dot{\gamma}^*(\bar{t}-s))^2
\right ].
\label{ED:def}
 \end{equation}
Hence, we obtain the time correlation of $\hat{P}(\bv{q},t)$
as
\begin{eqnarray}
\left < \hat{P}(\bv{q},\bar{t}) \hat{P}(\bv{q}',0) \right >
& = & \left < \tilde{P}(\bv{k},\bar{t}) \tilde{P}(\bv{k}',0) \right > \nonumber \\
& = & E_D(\bv{k},\bar{t}) \delta_{\bv{k}',-\tilde{\bv{k}}(\dot{\gamma}^*\bar{t})},
\label{P:cor}
 \end{eqnarray}
where we have used
$\left < \tilde{P}(\bv{k},0) \tilde{P}(\bv{k}',0) \right > =
\delta_{\bv{k}',-\bv{k}(\dot{\gamma}^*\bar{t})}$.
Then, 
substituting $\delta \hat{u}_\alpha(\bv{q},t) =  (t_E^{-1} \sigma^4) \tilde{z}_{\alpha +1}
(\bv{k},\bar{t})$ into Eq. \eqref{CS:eq}
with Eqs. \eqref{P:cor} and \eqref{Z:cor}, we obtain
\begin{eqnarray}
C_{{\rm S},\alpha \alpha}(t) & = & t_E^{-2} \sigma^2
\int \frac{d \bv{k}}{(2\pi)^3}\int \frac{d \bv{k}'}{(2\pi)^3}\nonumber \\
& & \langle  \tilde{z}_\alpha(\bv{k},\bar{t}) \tilde{z}_\alpha(\bv{k}',0)\rangle
 E_D(\bv{k},\bar{t}) \delta_{\bv{k}',-\bv{k}(\dot{\gamma}^*\bar{t})}, \nonumber \\
& = &
t_E^{-2} \sigma^2 \int \frac{d \bv{k}}{(2\pi)^3}
C_{\alpha+1 \, \alpha+1}(\bv{k},t) E_D(\bv{k},\bar{t}) ,
\label{CS:trans}
\end{eqnarray}
Substituting Eq. \eqref{C:ab}
into Eq. \eqref{CS:trans} with the aid of Eqs. \eqref{psit:def} and 
\eqref{ED:def},
we obtain
\begin{eqnarray}
C_{{\rm S}, \alpha \alpha}(t) & = & t_E^{-2} \sigma^2  \sum_{l,m=1}^4
C^{(lm)}_{{\rm S}, \alpha \alpha}(t)
\label{CS:exact}
\end{eqnarray}
with
\begin{eqnarray}
C^{(lm)}_{{\rm S}, \alpha \alpha}(t) & = &
\int \frac{d \bv{k}}{(2\pi)^3}
\int _0^\infty d\tau E^{(l)}(\bv{k},t+\tau) 
E^{(m)}(-\tilde{\bv{k}}(\dot{\gamma}^*\bar{t}),\tau)
\nonumber \\
& & 
\psi_{\alpha+1}^{(l)}(\bv{k})
\psi_{\alpha+1}^{(m)}(-\tilde{\bv{k}}(\dot{\gamma}^*\bar{t}))
F^{(lm)}( \tilde{\bv{k}}(\dot{\gamma}^*(\bar{t}+\tau)) ) \nonumber \\
& & E_D(\bv{k},\bar{t}).
\label{CSa:exact}
\end{eqnarray}
It should be noted that
$C^{(lm)}_{{\rm S}, \alpha \alpha}(t)$ for $l,m=1,2$
is the correlation of the longitudinal components of the velocity,
while
$C^{(lm)}_{{\rm S}, \alpha \alpha}(t)$ for $l,m=3,4$
is the correlation of the transverse components.
$C^{(lm)}_{{\rm S}, \alpha \alpha}(t)$ for other cases
is zero because $F^{(lm)}$ in Eq. \eqref{CSa:exact}
is zero as shown in Eq. \eqref{F:other}.

It should be noted that 
Eq. \eqref{CS:exact} describes
the velocity autocorrelation functions
for the sheared granular fluids for all-time region.
In the short-time regime $t < \dot{\gamma}^{-1}$
$C_{{\rm S}, \alpha \alpha}(t)$ in Eq. \eqref{CS:exact}, 
is reduced to the known
result for the fluid at equilibrium
$C_{{\rm S}, \alpha \alpha}(t) \sim t^{-3/2}$.
For $t>\dot\gamma^{-1}$,
they decay faster than $t^{-3/2}$.
The detailed calculation of the time correlation function 
in the long-time region is summarized in Appendix \ref{app:long_tail}.
Here, we only present the results of our calculation in this subsection.
  
The longitudinal components of the velocity autocorrelations
$C^{(lm)}_{{\rm S}, \alpha \alpha}(t)$ 
with $l,m=1,2$
approximately satisfy
\begin{eqnarray}
 |C^{(11)}_{{\rm S}, \alpha \alpha}(t)| &=& |C^{(22)}_{{\rm S}, \alpha \alpha}(t)|
= |C^{(12)}_{{\rm S}, \alpha \alpha}(t)|= |C^{(21)}_{{\rm S}, \alpha \alpha}(t)|  \nonumber \\
 \nonumber\\
&=&O((\dot \gamma t^{5/2})^{-1}), 
\label{C11}
\end{eqnarray}
See  \eqref{C11x}--\eqref{C12} in Appendix \ref{longitudinal:app} 
for the explicit expressions of 
$C^{(lm)}_{{\rm S}, \alpha \alpha}(t)$ 
with $l,m=1,2$.

The transverse modes of velocity autocorrelations
$C^{(lm)}_{{\rm S}, \alpha \alpha}(t)$ 
with $l,m=3,4$ 
are approximately given by
\begin{equation}
 C^{(33)}_{{\rm S}, x x}(t)\sim C^{(44)}_{{\rm S}, x x}(t)
\sim C^{(34)}_{{\rm S}, x x}(t)
\sim C^{(43)}_{{\rm S}, x x}(t)
 \propto 
(\dot \gamma t^{5/2})^{-1},
\label{Cxx}
\end{equation}
\begin{eqnarray}
 C^{(33)}_{{\rm S}, y y}(t)
 & \propto &
(\dot \gamma t^{5/2})^{-1}, \nonumber \\
 C^{(44)}_{{\rm S}, y y}(t)
& =& C^{(34)}_{{\rm S}, y y}(t)
=C^{(43)}_{{\rm S}, y y}(t) = 0
\label{Cyy}
\end{eqnarray}
\begin{eqnarray}
 C^{(33)}_{{\rm S}, z z}(t)
 & \propto &
(\dot \gamma t^{5/2})^{-1}, \nonumber \\
 C^{(34)}_{{\rm S}, z z}(t)
 & \sim &C^{(43)}_{{\rm S}, y y}(t) \sim
(\dot \gamma t^{7/2})^{-1}, \nonumber \\
 C^{(44)}_{{\rm S}, z z}(t)
 & \propto &
(\dot \gamma t^{9/2})^{-1}, 
\label{Czz}
\end{eqnarray}
The explicit expressions for Eqs. 
\eqref{Cxx}--\eqref{Czz} are given by
Eqs.  \eqref{C33x}, \eqref{C33z} in Appendix \ref{transverse:app}

Substituting 
Eqs. \eqref{C11}--\eqref{Czz} into Eq. \eqref{CS:exact},
the long-time behaviors of the velocity autocorrelation functions are given by
\begin{equation}
C_{{\rm S}, \alpha \alpha}(t)
\propto (\dot \gamma t^{5/2})^{-1}.
\label{C:longtime}
\end{equation}

We should note that the long-time behavior of $C_{{\rm S}, \alpha \alpha}(t)$
is previously obtained in Ref. \cite{otsuki07},
where the mixing term ${\sf L}_1$ in Eq. \eqref{matrix} is ignored 
with the method for the kinetic equations for dilute granular gases relying  on \cite{ernst71}.
The exponent for the long-time tail in Eq. \eqref{C:longtime}
is identical to that in Ref. \cite{otsuki07},
but the amplitude for $C_{{\rm S}, \alpha \alpha}(t)$ should be different.
The validity of the theoretical prediction for $C_{{\rm S}, \alpha \alpha}(t)$
has already been  verified by the numerical simulations with the introduction of one fitting parameter \cite{otsuki07}.
The results of the two-dimensional simulation  reasonably agree with the theoretical predictions, that is, (i) the more precise
treatment mentioned in this paper gives us only corrections of amplitude of the autocorrelation function, and (ii) the previous
results can be used for dense granular systems.  
If we ignore the second term on the right hand side of 
\eqref{lam1}--\eqref{lam3}  
which represents the mixing effect, 
$C_{{\rm S}, \alpha \alpha}(t)$
is reduced to that in Ref. \cite{otsuki07}.

Our prediction $t^{-5/2}$ is different from $t^{-9/2}$ \cite{Kumaran06,Kumaran09}.
We also stress that our theory gives reasonable agreement with the simulation
including the existence of anistropy in VACF
and the cross-over regime around $t\sim \dot\gamma^{-1}$ 
for two-dimensional case\cite{otsuki07}. Therefore, we believe that the velocity autocorrelation function 
satisfies $t^{-5/2}$ for $t\gg \dot\gamma^{-1}$, but the direct check of its validity based on simulation for three dimensional systems is difficult.

\subsection{long-range correlation function}

Our formulation can be used to discuss the equal-time spatial correlation functions.
It is obvious that the formulation of our theory is almost the same as that in Ref. \cite{Otsuki09},
and the results is identical to those reported in the previous paper.
Therefore, we only present the result of our analysis in this paper.

Since $\delta \hat{n}(\bv{q},t) =  \tilde{z}_1(\bv{k},\bar{t})$ and
$\delta \hat{u}_\alpha(\bv{q},t) =  t_E^{-1} \sigma^4 \tilde{z}_{\alpha +1}
(\bv{k},\bar{t})$,
the spatial correlations in Eqs. \eqref{Cnn:rew} and \eqref{Cpp:rew} are given by
\begin{eqnarray}
C_{nn}(\bv{r}) & = & \sigma^{-6}
\int \frac{d\bv{k}}{(2\pi)^3} \frac{d\bv{k}'}{(2\pi)^3} 
\langle \tilde{z}_1(\bv{k},0) 
\delta  \tilde{z}_1 (\bv{k}',0)\rangle \nonumber \\
& & e^{-i\{ \bv{k} \cdot \bv{r} +(\bv{k}+\bv{k}') \cdot \bv{r}' \} /\sigma},  \label{Cnn:z}\\
C_{pp}(\bv{r}) & = & (m n_0 \sigma t_E^{-1})^2 
\int \frac{d\bv{k}}{(2\pi)^3} \frac{d\bv{k}'}{(2\pi)^3}  \nonumber \\
& & \sum_{\alpha = 2,3,4}
\langle \tilde{z}_\alpha(\bv{k},0)  
\tilde{z}_\alpha (\bv{k}',0) \rangle \nonumber \\
& &
 e^{-i \{ \bv{k} \cdot \bv{r}/\sigma +(\bv{k}+\bv{k}') \cdot \bv{r}'\} /\sigma}.
\label{Cpp:z}
\end{eqnarray}

Substituting Eq. (\ref{Z:cor}) into Eqs. (\ref{Cnn:z})
and  (\ref{Cpp:z}), we obtain
\begin{equation}
C_{nn}(\bv{r}) = \sigma^{-6}
\int \frac{d\bv{k}}{(2\pi)^3}
C_{11}(\bv{k},0) e^{-i\bv{k} \cdot \bv{r}/\sigma},  
\label{Cnn:final}
\end{equation}
\begin{equation}
C_{pp}(\bv{r})  =  (m n_0 \sigma t_E^{-1})^2 \sum_{\alpha = 2,3,4}
\int \frac{d\bv{k}}{(2\pi)^3}
C_{\alpha \alpha}(\bv{k},0) 
 e^{-i\bv{k} \cdot \bv{r}/\sigma},
\label{Cpp:final}
\end{equation}
where the definition of $C_{\alpha \beta}(\bv{k},t)$ is given by Eq. (\ref{C:ab}).
As shown in  Ref. \cite{Otsuki09},
using the similar scaling procedure to obtain Eq. (\ref{C:longtime}),
the asymptotic behaviors of the spatial correlations
for
Eqs. (\ref{Cnn:final}) and (\ref{Cpp:final})
are given by
\begin{eqnarray}
C_{nn}(\bv{r}) & \propto & \left ( \frac{r}{l_c} \right )^{-11/3},
\qquad r \gg l_c, \\
C_{pp}(\bv{r}) & \propto & \left ( \frac{r}{l_c} \right )^{-5/3},
\qquad r \gg l_c, \label{Cpp:asymptotic}
\end{eqnarray}
which indicate the existence of the algebraic correlation
in $C_{nn}(\bv{r})$ and  $C_{pp}(\bv{r})$ with 
$l_c \equiv \sigma / \dot \gamma^*$.
The validity of the existence of long-range correlation
has been verified by the numerical simulation in  Ref. \cite{Otsuki09},
where the theory can be used, at least, until $\phi\le 0.50$ with the volume fraction $\phi$.

\section{Discussion and conclusion}
\label{Dis:sec}

 In this paper, we demonstrate that an unified description of both equal-time spatial correlation functions and the time correlation function 
for the uniformly sheared granular liquids with small inelasticity is possible
based on the generalized fluctuating hydrodynamics associated with FDR. 
The theory can be used for (i) an uniform sheared state with a constant temperature is stable, and (ii) 
FDR and Navier-Stokes like equation can be used. These assumptions are
justified for nearly elastic granular liquids in a small box under Lees-Edwards boundary condition. 
The quantitative validity of  our treatment has already been confirmed in Refs.  \cite{otsuki07} and \cite{Otsuki09}.
The theoretical prediction for $C_{{\rm S}, \alpha \alpha}(t)$ in this paper
is essentially the same as that in Ref. \cite{otsuki07} except for the amplitude.
Although the method in Ref. \cite{otsuki07} is only valid for dilute granular gases with small inelasticity, we now obtain more generalized method which can be used for
considerably dense granular gases. Indeed, semi-quantitative validity of our method for the equal-time spatial correlation functions has been confirmed
for $\phi\le 0.50$ \cite{Otsuki09}.  We also note that the method in Ref. \cite{otsuki07} ignore the mixing terms among transverse velocity mode,
the longitudinal mode and the density mode, but we now include such mixing effect in our calculation.
Thus, we conclude that  the contribution of mixing terms is small.

Although we have fixed the temperature in our analysis,
the temperature fluctuates in granular liquids.
However, if we take into account the fluctuation of the temperature,
the asymptotic behaviors of the correlations does not change.
Indeed, in Ref. \cite{Otsuki09r}, 
we have analyzed the long-ranged correlation function 
from the fluctuating hydrodynamics with a local transport law,
where the temperature fluctuation is considered,
and demonstrated that the asymptotic decay of the correlation is the same as
Eq. \eqref{Cpp:asymptotic}.

We have three directions to extend our work. One is to remove FDR in Eq. (\ref{noise})
to describe highly dissipative granular liquids,  another is to use soft-spheres to
describe very dense granular liquids including jamming transition and the other is to include the temperature fluctuation as well as nonlinear hydrodynamic equations
to describe non-uniform sheared granular liquids. The last one should be important to extract the characteristics of granular liquids under physical boundaries.

In conclusion, we demonstrate that an unified description of  both time correlation function and spatial correlation function is possible for 
uniformly sheared granular liquids.  This theory unifies our previous work written in Refs. \cite{otsuki07} and \cite{Otsuki09}. 
This method is also valid for sheared molecular gases controlled by the Gaussian thermostat. 
This unification may give us a simple view of sheared granular liquids.  

We thank S.-H. Chong for discussions. 
This work was supported by the Grant-in-Aid for scientific
research from the Ministry of Education, Culture, Sports,
Science and Technology (MEXT) of Japan 
(Nos.~21015016, 21540384, and 21540388),
by the Global COE Program
``The Next Generation of Physics, Spun from Universality
and Emergence'' from MEXT of Japan,
and in part by the Yukawa International Program for
Quark-Hadron Sciences at Yukawa Institute for Theoretical
Physics, Kyoto University.

\appendix

\section{The approximated expression of $C_{{\rm S}, \alpha \alpha}(t) $
by the velocity field}
\label{CS:app}

In this appendix, let us derive the approximated 
expression of $C_{{\rm S},\alpha\alpha}(t)$
given by Eq. (\ref{CS:eq}).
The derivation is based on the explanation in Ref. \cite{Zwanzig}. As a result, we can calculate $C_{{\rm S},\alpha\alpha}(t)$ by an unified method.

In order to obtain the approximated expression of 
$C_{{\rm S}, \alpha \alpha}(t) $, we introduce a complete orthonormal set of
functions $\phi_l(t)$ of the phase space point satisfying
$\langle \phi_l(t) \phi_m^*(t)\rangle = \delta_{lm}$.
Using $\phi_l(t)$, $C_{{\rm S}, \alpha \alpha}(t)$ in Eq. 
(\ref{C:def})
can be represented as
\begin{equation}
C_{{\rm S}, \alpha \alpha}(t) =  \sum_{l,m}
\langle  \delta v_{1,\alpha}(0) \phi_l^*(0) \rangle 
\langle   \phi_m(0) \delta v_{1,\alpha}(0) \rangle 
\langle  \phi_l(t) \phi_m^* \rangle,
\label{Projection:C}
\end{equation}
where we have used $\sum_{i} \langle \delta v_{i,\alpha}\delta v_{i,\alpha} \rangle
=N \langle \delta v_{1,\alpha}\delta v_{1,\alpha} \rangle$.

To extract the slow relaxation of $C_{{\rm S}, \alpha \alpha}(t) $,
we choose the slow variables
\begin{eqnarray}
\phi_{\bv{q},\alpha}(t) 
= \frac{1}{\sqrt{mNT}} \delta \hat{p}_{\alpha}(\bv{q},t) \hat{P}(-\bv{q}, t),
\label{slow:var}
\end{eqnarray}
where $\delta \hat{p}_{\alpha}(\bv{q},t)=m\sum_j \delta v_{j,\alpha}(t) e^{-i \bv{q}\cdot
\bv{r}_j(t)}$ and $\hat{P}({\bv{q}},t)=e^{-i \bv{q}\cdot \bv{r}_1(t)}$
are the Fourier components of the momentum density and the microscopic
concentration of the tagged particle, respectively.
We should note that 
$\langle  \delta v_{1,\alpha} \phi_{\bv{q},\beta}^* \rangle =\sqrt{T/(mN)}
\delta_{\alpha, \beta}$.
Then, substituting Eq. \eqref{slow:var} into Eq. \eqref{Projection:C},
we obtain
\begin{eqnarray}
C_{{\rm S}, \alpha \alpha}(t) 
& \simeq  & \sum_{\bv{q},\bv{q}'}
\langle  \delta v_{1,\alpha}(0) \phi_{\alpha}^*(\bv{q},0) \rangle 
\langle   \phi_{\alpha}(\bv{q}',0) \delta v_{1,\alpha}(0) \rangle 
\nonumber \\
& &
\langle  \phi_{\alpha}(\bv{q},t) \phi_{\alpha}^*(\bv{q}',0) \rangle,
\nonumber \\
& \simeq  & \frac{1}{V^2} \sum_{\bv{q},\bv{q}'}
\langle  \delta \hat{u}_{\alpha}(\bv{q},t) 
\delta \hat{u}_{\alpha}(\bv{q}',0) \rangle \nonumber \\
& & 
\langle \hat{P}(\bv{q},t) \hat{P}(\bv{q}',0) \rangle,
\end{eqnarray}
where we have used 
$\delta \hat{p}_{\alpha}(\bv{q},t) \simeq
mn_0 \delta \hat{u}_{\alpha}(\bv{q},t)$ with $n_0=N/V$,
$ \langle \hat{P}(-\bv{q},t) \hat{P}(-\bv{q}',0) \rangle
=\langle \hat{P}(\bv{q},t) \hat{P}(\bv{q}',0) \rangle$,
the decoupling approximation, and replaced $\bv{q}'$ by $-\bv{q}'$.
Finally, replacing the sum over $\bv{q}$ by an integral
as $\sum_{\bv{q}} \rightarrow V \int d\bv{q}/(2\pi)^3$,
we obtain the approximated expression of
$C_{{\rm S}, \alpha \alpha}(t)$
in Eq. \eqref{CS:eq}.

\section{Eigenvectors and eigenvalues}
\label{eigen:app}

In this appendix, let us explicitly write eigenvectors and eigenvalues for the eigenequation \eqref{eigen:eq} 
derived from the linearized fluctuating hydrodynamics. 

In order to obtain 
$\bv{\psi}^{(j)}(\bv{k})$, 
$\bv{\varphi}^{(j)}(\bv{k})$, 
and $\lambda^{(j)}(\bv{k})$,
we use the expansions
\begin{eqnarray}
\bv{\psi}^{(j)} & = & \bv{\psi}^{(j)}_{0} + \dot{\gamma}^* \bv{\psi}^{(j)}_{1} + \cdots, \label{expand1}\\
\bv{\varphi}^{(j)} & = &\bv{\varphi}^{(j)}_{0} + \dot{\gamma}^* \bv{\varphi}^{(j)}_{1} + \cdots, \label{expand2}\\
\lambda^{(j)} & = & \lambda^{(j)}_0 + \dot{\gamma}^* \lambda^{(j)}_1 + \cdots,\label{expand3}
\end{eqnarray}  
in terms of $\dot{\gamma}^*$.

We should note that the perturbation in terms of $\dot\gamma^*$ is not the expansion from an unsheared state
of granular liquids. Indeed, it is well-known that properties of sheared granular liquids completely differ
from those of freely cooling granular liquids. In the case of sheared granular liquids,
we obtain the relation
$\dot{\gamma}^* \sim \dot{\gamma} / \sqrt{T}
\sim \sqrt{1-e^2}$
from the balance between the viscous heating and the collisional energy loss. 
Thus, the expansion in terms of $\dot\gamma^*$ can be regarded as that 
by small inelasticity \cite{Alam08}.

Substituting Eqs. (\ref{expand1}), (\ref{expand2}) and (\ref{expand3})
into Eq. (\ref{eigen:eq}),
we obtain the zeroth and the first order perturbations as
\begin{eqnarray}
& & ( {\sf L}_{0} - \lambda^{(j)}_0 {\sf 1})\cdot
\bv{\psi}^{(j)}_{0}(\bv{k})  =  0, \\
& & ( {\sf L}_{0} - \lambda^{(j)}_0 {\sf 1})\cdot
\bv{\psi}^{(j)}_{1}(\bv{k}) \nonumber \\
& & \quad + \left( -  {\sf 1}k_x \frac{\partial}{\partial k_y}  + {\sf L}_{1} - \lambda^{(j)}_1 {\sf 1} \right)\cdot
\bv{\psi}^{(j)}_{0}(\bv{k}) 
 =  0.
\end{eqnarray}
Solving these equations, we obtain the eigenvalues given by 
Eqs. \eqref{lam1}--\eqref{lam4}.
Similarly,
we obtain
the right eigenvectors
\begin{eqnarray}
\bv{\psi}^{(1)T} & = & \frac{1}{N_+}\left(ikn_0 \sigma^3,\lambda_+ \frac{k_x}{k},\lambda_+ \frac{k_y}{k},\lambda_+ \frac{k_z}{k} \right), 
\label{psi1}\\
\bv{\psi}^{(2)T} & = & \frac{1}{N_-}\left(ikn_0 \sigma^3,\lambda_- \frac{k_x}{k},\lambda_- \frac{k_y}{k},\lambda_- \frac{k_z}{k} \right), \label{psi2}\\
\bv{\psi}^{(3)} & = & \bv{\Psi}^{(3)} + M(\bv{k})\bv{\Psi}^{(4)}, \label{psi3}\\
\bv{\psi}^{(4)} & = & \bv{\Psi}^{(4)},
\end{eqnarray}
and the left eigenvectors
\begin{eqnarray}
\label{phi^1}
\bv{\varphi}^{(1)} & = & \frac{1}{N_+}\left(ikp^*(k,e),\lambda_+ \frac{k_x}{k},\lambda_+ \frac{k_y}{k},\lambda_+ \frac{k_z}{k} \right), \label{phi1}\\
\bv{\varphi}^{(2)} & = & \frac{1}{N_-}\left(ikp^*(k,e),\lambda_- \frac{k_x}{k},\lambda_- \frac{k_y}{k},\lambda_- \frac{k_z}{k} \right), \label{phi2}\\
\bv{\varphi}^{(3)} & = & \bv{\Phi}^{(3)}, \\
\bv{\varphi}^{(4)} & = & -M(\bv{k})\bv{\Phi}^{(3)}+\bv{\Phi}^{(4)}, \label{phi^4}
\end{eqnarray}
where we have introduced 
\begin{eqnarray}
\bv{\Psi}^{(3)T} & = & \bv{\Phi}^{(3)} \equiv \left(0,-\frac{k_y k_x }{kk_\perp}, \frac{k_\perp}{k},-\frac{k_y k_z }{kk_\perp} \right) , \\
\bv{\Psi}^{(4)T} & = & \bv{\Phi}^{(4)}\equiv  \left(0, -\frac{k_z}{k_\perp},0,\frac{k_x}{k_\perp}\right),
\end{eqnarray}
and
\begin{equation}
M(\bv{k}) = - \frac{kk_z}{k_xk_\perp} \tan^{-1}(k_y/k_\perp) 
\label{M:def}
\end{equation}
with $k_\perp \equiv k^2 - k_y^2$.
It should be noted that these eigenvectors with $M(\bv{k})$ 
is valid only when $k_x \neq 0$ \cite{Lutsko85}.

\section{Solution of Eq. \eqref{a:eq}}
\label{sol:app}

In this appendix, we derive the solution \eqref{a:sol}
of Eq. \eqref{a:eq}.
Introducing $\bv{Q}\equiv \tilde{\bv{k}}(\dot{\gamma}^*\bar{t}')$ and
transformations
 $(\bar{t}, \bv{k}) = (\bar{t}', 
\tilde{\bv{Q}}(-\dot{\gamma}^*\bar{t}'))$ with $\bar{t}'=\bar{t}$ and
$\tilde{\bv{Q}}(\bar{\tau}) \equiv (Q_x,Q_y + \bar{\tau} Q_x,Q_z)$,
Eq. \eqref{a:eq} is rewritten as
\begin{equation}
\left(\partial_{\bar{t}'} 
+ \lambda^{(j)}(\tilde{\bv{Q}}(-\dot{\gamma}^*\bar{t}'))) \right) 
\tilde{A}^{(j)}(\bv{Q},\bar{t}')
 = F^{(j)}(\tilde{\bv{Q}}(-\dot{\gamma}^*\bar{t}')),\bar{t}'), \label{a:eqR}
 \end{equation}
 where we have introduced
$\tilde{A}^{(j)}(\bv{Q},\bar{t}') = \tilde{a}^{(j)}
(\tilde{\bv{Q}}(-\dot{\gamma}^*\bar{t}'),\bar{t}')$, and
used the relations 
$ \partial_{\bar{t}} =  \partial_{\bar{t}'} +
\dot{\gamma}^* Q_x \partial_{Q_y}$.

The solution of Eq. \eqref{a:eqR} is obtained as 
\begin{eqnarray}
\tilde{A}^{(j)}(\bv{Q},\bar{t}') & = &
\int_{\bar{t}'_0}^{\bar{t}'} ds
F^{(j)}(\tilde{\bv{Q}}(-\dot{\gamma}^*s),s)
\frac{E^{'(j)}(\bv{Q},\bar{t}')}{E^{'(j)}(\bv{Q},s)}
 \nonumber \\
 & &
+  
\tilde{A}^{(j)}(\bv{Q},\bar{t}'_0)
\frac{E^{'(j)}(\bv{Q},\bar{t}')}{E^{'(j)}(\bv{Q},\bar{t}'_0)},
 \end{eqnarray}
where $\bar{t}'_0$ is an initial time, and 
$E^{'(j)}(\bv{Q},\bar{t})$ is defined by 
\begin{eqnarray}
E^{'(j)}(\bv{Q},\bar{t})
&\equiv&
\exp[  - \int_0^{\bar{t}} d\bar{\tau} \lambda^{(j)}( \tilde{\bv{Q}}
(-\dot{\gamma}^*\bar{\tau}))], \label{Ed:def}
\end{eqnarray}
Taking the limit $\bar{t}'_0 \rightarrow - \infty$,
we obtain
\begin{equation}
\tilde{A}^{(j)}(\bv{Q},\bar{t}')  = 
\int_{-\infty}^{\bar{t}'} ds
F^{(j)}(\tilde{\bv{Q}}(-\dot{\gamma}^*s),s)
\frac{E^{'(j)}(\bv{Q},\bar{t}')}{E^{'(j)}(\bv{Q},s)}.
 \end{equation}
Since 
$\tilde{a}^{(j)}(\bv{k},\bar{t}) = \tilde{A}^{(j)}
(\tilde{\bv{k}}(\dot{\gamma}^*\bar{t}),\bar{t}')$,
\begin{equation}
\tilde{a}^{(j)}(\bv{k},\bar{t})  = 
\int_{-\infty}^{\bar{t}} ds
F^{(j)}(\tilde{\bv{k}}(\dot{\gamma}^*(\bar{t}-s)),s)
\frac{E^{'(j)}(\tilde{\bv{k}}(\dot{\gamma}^*\bar{t}),\bar{t})}
{E^{'(j)}(\tilde{\bv{k}}(\dot{\gamma}^*\bar{t}),s)}.
\label{ap:eq}
 \end{equation}

Here, from the transformation $ \bar{\tau}' = \bar{t}-\bar{\tau}$,
we obtain
\begin{eqnarray}
\frac{E^{'(j)}(\tilde{\bv{k}}(\dot{\gamma}^*\bar{t}),\bar{t})}
{E^{'(j)}(\tilde{\bv{k}}(\dot{\gamma}^*\bar{t}),s)}
& =  &
\frac{\exp[  - \int_0^{\bar{t}} d\bar{\tau} \lambda^{(j)}
( \tilde{\bv{k}}(\dot{\gamma}^*(\bar{t}-\bar{\tau})))]}
{\exp[  - \int_0^{s} d\bar{\tau} \lambda^{(j)}( \tilde{\bv{k}}(\dot{\gamma}^*(\bar{t}-\bar{\tau})))]} \nonumber \\
&= & 
\frac{\exp[  - \int_0^{\bar{t}} d\bar{\tau}' \lambda^{(j)}
( \tilde{\bv{k}}(\dot{\gamma}^*\bar{\tau}'))]}
{\exp[  - \int_{\bar{t}-s}^{\bar{t}} d\bar{\tau}' 
\lambda^{(j)}( \tilde{\bv{k}}(\dot{\gamma}^*\bar{\tau}'))]}
\nonumber \\
&= & 
\exp[  - \int_0^{\bar{t}-s} d\bar{\tau}' \lambda^{(j)}
( \tilde{\bv{k}}(\dot{\gamma}^*\bar{\tau}'))] \nonumber \\
& = & E^{(j)}(\tilde{\bv{k}},\bar{t}-s),
\end{eqnarray}
where 
$E^{(j)}(\bv{k},\bar{t})$ is given by Eq. \eqref{E:def}.
Substituting this equation into Eq. \eqref{ap:eq},
we obtain Eq. \eqref{a:sol}.

\section{Long-time behaviors of the time correlation functions}
\label{app:long_tail}

In this appendix, we explicitly evaluate the long-time behaviors of
the time correlation functions given in Sec. \ref{long_time:sec}.
In the first part, we obtain the general expressions
for the long-time behaviors of the time correlation functions.
In the second part, we evaluate the long-time behaviors of the time correlations
of the longitudinal component of the velocity.
In the third part, the time correlations
of the transverse component of the velocity is estimated.

\subsection{The general expressions
for the long-time behaviors of the time correlation functions}

Introducing $\bv{K}$ as $K_x\equiv k_x \dot\gamma^* \bar{t}^{3/2}$ and 
$K_{\alpha}\equiv k_{\alpha} \bar{t}^{1/2}$ for $\alpha\ne x$,
and $\tau' = \tau/t$,
Eq. \eqref{CSa:exact} are replaced by
\begin{eqnarray}
C^{(lm)}_{{\rm S}, \alpha \alpha}(t) & = & (\dot \gamma^* \bar{t}^{3/2})^{-1}
\int \frac{d \bv{K}}{(2\pi)^3}
\int _0^\infty d\tau'  \nonumber \\
& &  F^{(lm)}( \tilde{\bv{K}}+\bar{t}^{-1/2}(1+\tau')K_x)
\nonumber \\
& & E^{(l)}(\tilde{\bv{K}}
,\bar{t}(1+\tau'))  
 E^{(m)}(-\tilde{\bv{K}}-\bar{t}^{-1/2}\bv{e}_yK_x, \bar{t} \tau')
\nonumber \\
& & 
\psi_{\alpha+1}^{(l)}(\tilde{\bv{K}})
\psi_{\alpha+1}^{(m)}(-\tilde{\bv{K}}-\bar{t}^{-1/2}\bv{e}_y K_x)
\nonumber \\
& & 
E_D(\tilde{\bv{K}},\bar{t}).
\label{CSaa}
\end{eqnarray}
where we have introduced $\tilde{\bv{K}} \equiv \bar{t}^{-1/2}(K_x/(\dot\gamma^* \bar{t}),K_y,K_z)$.

In the long-time regime $t \gg \dot \gamma^{-1}$, 
since $\tilde{\bv{K}} \rightarrow 0$,
$E^{(l)}(\tilde{\bv{K}},\bar{t}\tau') $ and $E_D(\tilde{\bv{K}},\bar{t}\tau') $
are approximately given  by
\begin{eqnarray}
& & |E^{(1)}(\tilde{\bv{K}},\bar{t} \tau') | \nonumber \\
 & &  \quad \simeq \exp \left [ -\frac{1}{2} \int_0^{\tau'} ds' \nu^*_1(0,e)
 K_T(s')^2 \right ] 
\sqrt{\frac{K_T}{K_T(\tau')}}, 
\label{E1} \\
&  & |E^{(2)}(\tilde{\bv{K}},\bar{t} \tau') 
= |E^{(1)}(\tilde{\bv{K}},\bar{t}\tau') |, 
\end{eqnarray}
\begin{eqnarray}
& & E^{(3)}(\tilde{\bv{K}},\bar{t}\tau')  \nonumber \\
& & \quad \simeq  \exp \left [ -\int_0^{\tau'} ds' \nu^*_2(0,e)
K_T(s')^2  \right ]\frac{K_T(\tau')}{K_T}, \\
& & E^{(4)}(\tilde{\bv{K}},\bar{t}\tau')  
 = E^{(3)}(\tilde{\bv{K}},\bar{t}\tau')\frac{K_T}{K_T(\tau')}, \nonumber \\
& & E_D(\tilde{\bv{K}},\bar{t}) 
\simeq  \exp \left [ -\int_0^{1} ds' D^* K_T(s')^2 \right ],
\end{eqnarray}
where we have introduced $K_T^2(\tau) = (K_y + \tau K_x)^2 + K_z^2$,
and $K_T^2$ implies $K_T(0)$.

$\bv{\psi}^{(l)}(\tilde{\bv{K}})$
in the long-time regime is approximated by
\begin{eqnarray}
& & \bv{\psi}^{(1)T}(\tilde{\bv{K}}) 
 \simeq  
\frac{1}{\sqrt{2}} \left \{ \bv{\Psi}_0+\bv{\Psi}_L \right \}, \\
& & \bv{\psi}^{(2)T}(\tilde{\bv{K}}) 
\simeq  
\frac{1}{\sqrt{2}} \left \{ \bv{\Psi}_0-\bv{\Psi}_L \right \}, \\
& & \bv{\psi}^{(3)T}(\tilde{\bv{K}}) \nonumber \\ 
& & \simeq  
\left (0, -M'(\bv{K}) \frac{K_z}{ |K_z|},
\frac{|K_z|}{K_{\rm T}},
-\frac{K_y K_z}{K_{\rm T}|K_z|} \right ) \\
& & \bv{\psi}^{(4)T}(\tilde{\bv{K}})
 \simeq  \left (0, - \frac{K_z}{ |K_z|},0,\frac{K_x}{ \dot \gamma^* \bar{t} |K_z|} \right )
\end{eqnarray}
with
\begin{eqnarray}
\bv{\Psi}_0  & \simeq  &
\left(1,0,0,0 \right), \\
\bv{\Psi}_L  & \simeq  &
\left(0,
\frac{K_x}{(\dot \gamma^* \bar{t}) K_{\rm T}},
\frac{K_y}{K_{\rm T}},
\frac{K_z}{K_{\rm T}}, \right),
\end{eqnarray}
where we have introduced 
$M'(\bv{K}) = \frac{K_{\rm T} K_z}{\Lambda |K_z|} \tan^{-1}\left (
\frac{K_y}{|K_z|}
\right )$.
Here, we have introduced the infrared cut off $\Lambda$ for $k_x$
to avoid the divergence of $M'(\bv{K})$.

$F^{(lm)}(\tilde{\bv{K}})$
in the long-time regime is approximated by
\begin{eqnarray}
& & F^{(11)}(\tilde{\bv{K}}) 
 \simeq  -\frac{A \nu_1^*(0,e) K_T^2}{ \bar{t}}, \\
& & F^{(22)}  = F^{(12)}=  F^{(21)}=  F^{(11)} ,
\end{eqnarray}
\begin{eqnarray}
& & F^{(33)} (\tilde{\bv{K}}) 
 \simeq \frac{2A \nu_2^*(0,e) K_T^2}{ \bar{t}}, \\
& & F^{(44)}(\tilde{\bv{K}}) 
\simeq  
 - N(\bv{K})F^{(33)}(\tilde{\bv{K}}), \\
& & F^{(34)}(\tilde{\bv{K}}) 
 = -  F^{(43)}(\tilde{\bv{K}}) 
\simeq  M'(\bv{K}) F^{(33)}(\tilde{\bv{K}})
 \label{F43}
\end{eqnarray}
with $N(\bv{K})= M'(\bv{K})^2 + 1$.

\subsection{The time correlations for the longitudinal components
of the velocity fields}
\label{longitudinal:app}

Substituting Eqs. (\ref{E1})--(\ref{F43}) into Eq. (\ref{CSaa}),
$C^{(lm)}_{{\rm S}, \alpha \alpha}(t)$ for $l,m = 1,2$ 
in the long-time regime is approximately
given by
\begin{eqnarray}
|C^{(11)}_{{\rm S}, x x}(t)| & \le & \frac{A \nu_1^*(0,e) }{2}
(\dot \gamma^{*3} \bar{t}^{9/2})^{-1} \nonumber \\
& & \int \frac{d \bv{K}}{(2\pi)^3}
\int _0^\infty d\tau'   G^{(11)}_x(\bv{K},\tau'),
\label{C11x}
\end{eqnarray}
and
\begin{eqnarray}
|C^{(11)}_{{\rm S}, \alpha \alpha}(t)| & \le & \frac{A \nu_1^*(0,e) }{2}
(\dot \gamma^{*3} \bar{t}^{5/2})^{-1} \nonumber \\
& & \int \frac{d \bv{K}}{(2\pi)^3}
\int _0^\infty d\tau'   G^{(11)}_\alpha(\bv{K},\tau')
\end{eqnarray}
for $\alpha = y, z$,  and
\begin{equation}
|C^{(11)}_{{\rm S}, \alpha \alpha}(t)| = |C^{(22)}_{{\rm S}, u_\alpha u_\alpha}(t)|
= |C^{(12)}_{{\rm S}, \alpha \alpha}(t)|= |C^{(21)}_{{\rm S}, u_\alpha u_\alpha}(t)|,
\label{C12}
\end{equation}
where we have introduced
\begin{eqnarray}
 G^{(11)}_\alpha(\bv{K},\tau')
& = & 
H_1(\bv{K},\tau') \frac{K_T(1+\tau')K_\alpha^2}{\sqrt{K_TK_T(1)}},
\end{eqnarray}
with
\begin{eqnarray}
H_1(\bv{K},\tau') & = & 
\exp \left [ -\frac{1}{2} \int_0^{1+\tau'} ds' \nu^*_1(0,e)
K_T(s')^2 \right ] \nonumber \\
& & 
\exp \left [ -\frac{1}{2} \int_0^{\tau'} ds' \nu^*_1(0,e)
K_T(s')^2 \right ] \nonumber \\
& & \exp \left [ -\int_0^{1} ds' D^* K_T(s')^2 \right ].
\end{eqnarray}
Eqs. \eqref{C11x}--\eqref{C12} are the explicit expressions
for the long-time behaviors given by Eq. \eqref{C11}
in  Sec. \ref{long_time:sec}.

\subsection{The time correlations for the transverse components
of the velocity fields}
\label{transverse:app}

Substituting Eqs. (\ref{E1})--(\ref{F43}) into Eq. (\ref{CSaa}),
$C^{(lm)}_{{\rm S}, \alpha \alpha}(t)$ for $l,m = 3,4$ 
with $\alpha = x, y$ is given by
\begin{eqnarray}
C^{(lm)}_{{\rm S}, \alpha \alpha}(t) & \simeq & \frac{A \nu_1^*(0,e) }{2}
(\dot \gamma^{*3} \bar{t}^{5/2})^{-1} \nonumber \\
& & \int \frac{d \bv{K}}{(2\pi)^3}
\int _0^\infty d\tau'   G^{(lm)}_\alpha(\bv{K},\tau'),
\label{C33x}
\end{eqnarray}
where we have introduced
\begin{equation}
G^{(33)}_x(\bv{K},\tau')
 =
H_2(\bv{K},\tau')
\frac{K_T(1+\tau')^4}{\Lambda^2} T(\bv{K},0)T(\bv{K},1),
\end{equation}
\begin{equation}
G^{(33)}_y(\bv{K},\tau')
=  H_2(\bv{K},\tau')
 \frac{K_T(1+\tau')^4 K_z^2}
 {K_T^2K_T(1)^2},
\end{equation}
\begin{eqnarray}
G^{(44)}_x(\bv{K},\tau')
& \simeq & 
\left \{
 \frac{  K_T(1+\tau')^2  } {\Lambda^2} T(\bv{K},1+\tau')^2 + 1 \right \} 
 \nonumber \\
 & & H_2(\bv{K},\tau')K_T(1+\tau')^2,
\end{eqnarray}
\begin{eqnarray}
G^{(34)}_x(\bv{K},\tau')
& \simeq & -
 \frac{K_T(1+\tau')^4 }{\Lambda^2  }T(\bv{K},1+\tau')T(\bv{K},0) \nonumber \\
 & & H_2(\bv{K},\tau'),
\end{eqnarray}
\begin{eqnarray}
G^{(43)}_x(\bv{K},\tau')
& \simeq & -
 \frac{K_T(1+\tau')^4 }{\Lambda^2  }T(\bv{K},1+\tau')T(\bv{K},1) \nonumber \\
& &  H_2(\bv{K},\tau'),
\end{eqnarray}
\begin{equation}
G^{(44)}_{y}(\bv{K},\tau') =G^{(34)}_{y}(\bv{K},\tau') =G^{(43)}_{y}(\bv{K},\tau') = 0,
\end{equation}
with
\begin{eqnarray}
H_2(\bv{K},\tau') & = & 
\exp \left [ - \int_0^{1+\tau'} ds' \nu^*_2(0,e)
K_T(s')^2 \right ] \nonumber \\
& & 
\exp \left [ -\int_0^{\tau'} ds' \nu^*_2(0,e)
K_T(s')^2 \right ] \nonumber \\
& & \exp \left [ -\int_0^{1} ds' D^* K_T(s')^2 \right ],
\end{eqnarray}
and
\begin{eqnarray}
T(\bv{K},\tau) =
 \tan^{-1}\left ( \frac{K_y+\tau K_x}{|K_z|} \right ).
\end{eqnarray}
Eq. \eqref{C33x} is the explicit expression for 
long-time behaviors given by Eqs. \eqref{Cxx} and \eqref{Cyy}
in  Sec. \ref{long_time:sec}.

Similarly, from Eqs. (\ref{E1})--(\ref{F43}) into Eq. (\ref{CSaa}),
$C^{(lm)}_{{\rm S}, z z}(t)$ for $l,m = 3,4$ 
in the long-time regime is approximately
given by
\begin{eqnarray}
C^{(33)}_{{\rm S}, z z}(t) & \simeq & \frac{A \nu_1^*(0,e) }{2}
(\dot \gamma^{*} \bar{t}^{5/2})^{-1} \nonumber \\
& & \int \frac{d \bv{K}}{(2\pi)^3}
\int _0^\infty d\tau'   G^{(33)}_\alpha(\bv{K},\tau'), \nonumber \\
C^{(44)}_{{\rm S}, z z}(t) & \simeq & \frac{A \nu_1^*(0,e) }{2}
(\dot \gamma^{*3} \bar{t}^{9/2})^{-1} \nonumber \\
& & \int \frac{d \bv{K}}{(2\pi)^3}
\int _0^\infty d\tau'   G^{(44)}_\alpha(\bv{K},\tau'), \nonumber \\
C^{(34)}_{{\rm S}, z z}(t) & \simeq & \frac{A \nu_1^*(0,e) }{2}
(\dot \gamma^{*2} \bar{t}^{7/2})^{-1} \nonumber \\
& & \int \frac{d \bv{K}}{(2\pi)^3}
\int _0^\infty d\tau'   G^{(34)}_\alpha(\bv{K},\tau'), \nonumber \\
C^{(43)}_{{\rm S}, z z}(t) & \simeq & \frac{A \nu_1^*(0,e) }{2}
(\dot \gamma^{*2} \bar{t}^{7/2})^{-1} \nonumber \\
& & \int \frac{d \bv{K}}{(2\pi)^3}
\int _0^\infty d\tau'   G^{(43)}_\alpha(\bv{K},\tau'), 
\label{C33z}
\end{eqnarray}
where we have introduced
\begin{equation}
 G^{(33)}_z(\bv{K},\tau')
  =  
 H_2(\bv{K},\tau')  
 \frac{K_T(1+\tau')^4 K_y(K_y+K_x)}
 {K_T^2K_T(1)^2},
\end{equation}
\begin{eqnarray}
 G^{(33)}_z(\bv{K},\tau')
 & = &
\left \{
 \frac{  K_T(1+\tau')^2  } {\Lambda^2} T(\bv{K},1+\tau')^2 + 1 \right \} \nonumber \\
& &  H_2(\bv{K},\tau')  \frac{ K_T(1+\tau')^2K_x^2}{K_z^2},
\end{eqnarray}
\begin{equation}
 G^{(34)}_z(\bv{K},\tau')
 = 
 H_2(\bv{K},\tau')
 \frac{ K_T(1+\tau')^2K_y K_x}{\Lambda K_T^2 |K_z|  }
 T(\bv{K},1+\tau'),
\end{equation}
\begin{equation}
 G^{(43)}_z(\bv{K},\tau')
=  
 H_2(\bv{K},\tau') 
 \frac{K_T(1+\tau')^4K_y K_x}{\Lambda K_T(1)^2 |K_z|  }
 T(\bv{K},1+\tau').
\end{equation}
Equation \eqref{C33z} is the explicit expression for 
long-time behaviors given by Eq. \eqref{Czz} in  Sec. \ref{long_time:sec}.

%

\end{document}